\documentclass[aps,prb,preprint,groupedaddress,showpacs,showkeys]{revtex4-1}

\usepackage{amsmath}
\usepackage{amssymb}
\usepackage{graphicx}
\usepackage{epstopdf}

\usepackage{subcaption}
\usepackage{mwe}

\usepackage{media9}
\usepackage{hyperref}

\begin{document}

\title{Static and dynamic properties of heavily doped quantum vortices}
\author{I.A. Pshenichnyuk
\email[correspondence address: ]{ivan.pshenichnyuk@gmail.com}}
\affiliation{Skolkovo Institute of Science and Technology Novaya St., 100, Skolkovo 143025, Russian Federation}
\date{\today}

\begin{abstract}
Quantum vortices in superfluids may capture matter and deposit it inside their core. By doping vortices with foreign particles one can effectively visualize them and study experimentally. To acquire a better understanding of the interaction of quantum vortices with matter and clarify the details of recent experiments properties of doped vortices are investigated here theoretically in the regimes where the doping mass becomes close to the total mass of superfluid particles forming a vortex. Such formations are dynamically stable and, possessing both vorticity and enhanced inertia, demonstrate properties which are different from the pure vortex case. The goal of this paper is to define and investigate the universal aspects of a heavily doped vortex behavior which can be realized in different types of quantum mixtures. The proposed 3D model is based on a system of coupled semiclassical matter wave equations which are solved numerically in a wide range of physical parameters. The size, geometry, solubility and binding energy of dopants in different regimes are discussed. A coupled motion of a vortex-dopant complex and decoupling conditions are studied. The reconnection of vortices, taken as an example of a fundamental process responsible for the evolution of a quantum turbulent state, is modeled to illustrate the difference between the light and heavy doping cases.
\end{abstract}      


\maketitle


\section{Introduction}
\label{sec_intro}

Interaction of quantum vortices with impurities in superfluids is a nontrivial process giving insight into fundamental physical questions. As it was discussed previously \cite{pshenichnyuk-2016}, the scattering of particles on vortices is an inelastic process, accompanied by the energy redistribution through the emission of Kelvin waves. If the energy of particles is low enough, they can be captured by a vortex, forming a stable complex, which is referred as a doped vortex in this paper. Being loaded with an additional mass, which can be high, depending on the type of dopants used, the vortex demonstrates properties obviously different from the undoped case.

In dense quantum fluids, like liquid helium, where the vortex core size is of the order of 1 \AA, doping is a widely used experimental technique which makes vortices visible for detectors.\cite{bewley-2006, bewley-2008}  Understanding of the doped vortex dynamics, thus, is important for the interpretation of the experimental data on quantum turbulence.
In ultracold atomic gases experiments, where multicomponent Bose-Einstein condensates (BEC) can be created and controlled with a great degree of accuracy via a Feshbach resonance \cite{stan-2004}, appearance of quantized vortices may lead to a peculiar way of matter organization, when a phase-separated fraction of one component is captured by a vortex formed in the other component \cite{matthews-1999, williams-1999, kasamatsu-2005}. The behavior of such composite formations represents a fundamental physical interest. Due to the fast development of the ultracold gases field and successful attempts to realize BEC on a chip concept \cite{hansel-2001}, quantum turbulence based phenomena are becoming more common and may take its place in potential applications of BEC in future electronics.
Another example which motivates the investigation of doped quantum vortices is connected with a metallic nanowires production technique based on a quantum turbulence \cite{gordon-2009, moroshkin-2010, gordon-2012, gordon-2015}. It is shown experimentally that the ablation of metals in superfluid helium with laser pulses leads to the formation of centimeter-long wires produced from atoms trapped by quantized vortex filaments. Complex elongated nanostructures with a core and a shell made from different materials also can be produced using quantum vortices based technique\cite{thaler-2014}.
All mentioned examples, both fundamental and applied, show the importance of in-depth understanding of particle-vortex scattering and doped vortex dynamics for the subsequent progress in this fields.

Illuminating experiments in quantum turbulence were performed recently using superfluid helium nanodroplets \cite{gomez-2012, spence-2014, gomez-2014, jones-2016}. Vortex filaments in rotating droplets were doped with Ag and Xe atoms and studied using a femtosecond x-ray coherent diffractive imaging technique as well as an electron microscopy preceded by a surface-deposition of the samples. Several questions were raised in this works, including the origin of unusual shapes of doped helium droplets and distribution of vortices inside the droplet. Although certain aspects were clarified by theorists \cite{ancilotto-2003, ancilotto-2014, ancilotto-2015}, the connection between the rotational motion of a droplet and a dopant still remains unclear \cite{jones-2016}.

In xenon doped nanodroplets experiments doping particles are approximately 33 times heavier than fluid atoms and the diameter of each particle is comparable with a vortex core size in helium. It is quite opposite, for instance, in comparison to large and light electron bubbles often used as dopants and well studied in the past both experimentally and theoretically \cite{yarmchuk-1979, clark-1965, berloff-2000}. It is known that a particle radius is closely linked with a particle-vortex binding energy \cite{parks-1966}, which is the main interaction defining parameter. It means that, on the one hand, we can expect a small xenon atom capture to be accompanied by less intensive Kelvin waves generation and less perturbations in a vortex geometry \cite{pshenichnyuk-2016}. On the other hand, being captured, heavy xenon may influence the vortex motion significantly and theoretical modeling is necessary to understand the details of its behavior.

According to the results of Gordon and colleagues \cite{gordon-2009,moroshkin-2010}, guest atoms in helium above critical temperature tend to form spherical clusters. In superfluid helium below critical temperature impurities with a certain probability stick together to produce long cylindrical filaments, which are attributed to the presence of quantum vortices. These filaments are stable enough to exist independently, by decoupling from vortices. In xenon doped nanodroplets, for instance, xenon-xenon binding energy (24 meV) is significantly larger than corresponding helium-xenon (3 meV) and helium-helium (1 meV) energies \cite{tang-2003}. It is also larger than xenon-vortex binding (about 0.2 meV for a single atom \cite{pshenichnyuk-2015}). It makes filaments quite stable after the decoupling and allows to study them using a surface deposition technique. The described behavior also reveals the difference between atomic dopants and electron bubbles and motivates the necessity of theoretical modeling for the better understanding of heavy vortex dynamics.

Taking the listed experiments as a motivation, the goal of this paper is to study theoretically the general case of quantum vortex interacting with a heavy matter, the case which may be realized not only in helium, but in different kinds of multicomponent quantum mixtures. The proposed classical matter fields based 3D model allows to treat such parameters of the doping substance as a mass, volume and degree of solubility. It can be used to describe the dynamics of the interaction including scattering, coupled motion and decoupling of doping matter.
The theoretical model is introduced in Sec.~\ref{sec_theory} of the paper, where the universal dimensionless equations and corresponding Hamiltonian are presented. The space of physically relevant parameters typical for atomic dopants (both heavy and relatively light) is also defined and discussed there. An example of the optimized stationary solution representing a heavy quantum vortex with a dopant trapped and distributed homogeneously along its core is shown.
In Sec.~\ref{sec_stat} the defined space of parameters is investigated to check the existence and stability of phase-separated solutions. The cases where doping matter is trapped by vortices and self-trapped (using the terminology of Gross \cite{grant-1974}) inside the superfluid bulk are considered. A special attention is payed to the geometry and full energy of solutions which is used to determine the most favorable configurations. Sizes of doping matter fractions and binding energies between them and quantum vortices are calculated and discussed. The interval of masses covering the light to heavy doping transition is scanned to check the smoothness of the transition with respect to the properties of the system.
After defining a class of steady solutions of interest, the discussion continues in Sec.~\ref{sec_dynam}, where the dynamics of the mixture is probed. A coupled vortex-dopant propagation in a superfluid and decoupling conditions are investigated. The reconnection of two heavy vortices, as an example of a common phenomena in the turbulent regime, is modeled. The difference between the light and heavily vortices behavior is emphasized. The conclusion is given in Sec.~\ref{sec_conclusion}.


\section{Theory}
\label{sec_theory}

To build a model of a heavy quantum vortex the classical field formalism is applied \cite{svistunov}. The nonlinear field equation, also known as the generalized nonlinear Schrodinger equation (gNLSE) \cite{berloff-2014}, has vortical solutions (along with a family of solitary wave solutions) and presents the most complete mathematical model of a quantum vortex behavior. The proposed framework is quite universal. It allows to take into account an equation of state of a superfluid by adding a realistic internal energy functional into the Hamiltonian \cite{berloff-2009, berloff-2014}. An arbitrary form of nonlocality of particles interactions can be treated as well\cite{berloff-2000a}. It was demonstrated that gNLSE based formalism has a mathematical parallelism with the Landau two-fluid model and can be used to describe not only the condensed phase but also a cloud of thermal excitations, and thus applicable for the modeling of finite temperatures \cite{berloff-2002, berloff-2007}. In the context of liquid helium modeling it allows to describe both the superfluid and the normal fluid components \cite{berloff-2002, berloff-2014}.

In the present work the simplest form of the equation with a cubic nonlinearity and local interactions is used. The superfluid and the doping substance are modeled using a system of coupled nonlinear matter fields. In the present form the equations, being a good model for BEC mixtures\cite{tylutki-2016}, do not allow to model good enough the liquid helium, where only about 10\% of the fluid is in a condensed state. But they allow to track the universal mechanical aspects of the vortex-matter interaction, taking place in idealized quantum mixtures. Similar models were used in the past to understand qualitatively the behavior of electron bubbles in liquid helium \cite{grant-1974, berloff-2000}. Numerical results obtained here for the binding energy of particles to vortices, when mass and size parameters typical for xenon dopants in helium are used, correlate well with the results obtained using more sophisticated versions of the equation and DFT for helium (see below). Various extensions can be incorporated into the formalism, if necessary, to investigate the effects caused by the nonlocal interactions of particles and finite temperatures.

The Hamiltonian of a mixture of two incoherently interacting superfluids described by the complex valued classical fields $\psi$ and $\varphi$ reads
\begin{equation}
H = \int  \left\{ \frac{\hbar^2}{2m_1} |\nabla\psi|^2 + \frac{\hbar^2}{2m_2} |\nabla\varphi|^2 + g_{12} |\psi|^2|\varphi|^2 +\frac{g_{11}}{2}|\psi|^4 +\frac{g_{22}}{2}|\varphi|^4 - \mu_1|\psi|^2 - \mu_2 |\varphi|^2 \right\} dV , 
\end{equation}
where the integral is taken over the volume of the system $V$. Masses, chemical potentials and interparticle interactions are denoted as $m_i$, $\mu_i$ and $g_{ii}$ ($i = 1,2$). The interaction between two fluids is expressed through the parameter $g_{12}$. 
Corresponding equations of motion read
\begin{equation}
\begin{array}{c}
-i\hbar\psi_t = \frac{\hbar^2}{2m_1}\nabla^2\psi -g_{11}|\psi|^2\psi - g_{12}|\varphi|^2\psi + \mu_1\psi , \\
-i\hbar\varphi_t = \frac{\hbar^2}{2m_2} \nabla^2\varphi -g_{22}|\varphi|^2\varphi - g_{12}|\psi|^2\varphi + \mu_2\varphi .
\end{array}
\end{equation}
The fields $\psi$ and $\varphi$, which are associated with the fluid and the  doping substance, are assumed to be normalized
\begin{equation}
\begin{array}{c}
\int |{\psi}|^2 dV = N_1 , \,\,\,\,  \int |{\varphi}|^2 dV= N_2 , 
\end{array}
\end{equation}
where $N_1$ and $N_2$ are the numbers of particles in the fluid and doping substance respectively, and $N_2$ is assumed to be much smaller than $N_1$.
Chemical potentials $\mu_1$ and $\mu_2$ are connected with the amounts of particles $N_1$ and $N_2$ and chosen to let the homogenous solutions  $\psi_{\infty}=\sqrt{N_1/V}$ and $\varphi_{\infty}=\sqrt{N_2/V}$ to fulfill the decoupled ($g_{12}=0$) system of equations. It is easy to show that $\mu_1 = g_{11}\psi_{\infty}^2$ and $\mu_2 = g_{22}\varphi_{\infty}^2$.
Both components are assumed to be bosonic. In this work the accent is made on the modeling of heavy cores and mass effects, while probable particles statistics based phenomena are not considered. Possible models for fermionic liquids are described in the literature \cite{maruyama-2005, wen-2014, caballerobenitez-2009, tylutki-2016a}.

The interparticle interaction is assumed to be repulsive, which is the case for dense superfluids. In BEC experiments the repulsive interaction corresponds to one of the possible regimes, controlled via a Feshbach resonance. Interaction parameters $g_{11}$, $g_{22}$ and $g_{12}$ are connected with scattering lengths of particles as follows
\begin{equation}
g_{11} = \frac{4\pi{l_1}\hbar^2}{m_1}, \,\,\,
g_{22} = \frac{4\pi{l_2}\hbar^2}{m_2}, \,\,\,
g_{12} = \frac{2\pi{l_{12}}\hbar^2}{m_{12}} , 
\end{equation}
where $m_{12}=m_1m_2/(m_1+m_2)$ is the reduced mass and $l_{12} = l_1/2 + l_2/2$.
The model is completely defined by seven primary parameters: $m_i$, $l_i$, $N_i$ and $V$ ($i=1,2$).

A characteristic length scale in the system is associated with the first fluid (an amount of the second component is small). The healing length reads $\xi = \hbar / \sqrt{2m_1g_{11}\psi_{\infty}^2}$, or being expressed through the primary parameters $\xi = \sqrt{V/8\pi{l_1}N_1}$. The healing length is of the order of an angstrom for helium and it can be of the order of micrometers in BEC systems, which corresponds to an approximate size of vortex cores.

For practical computations the equations are transformed to the dimensionless form \cite{berloff-2005,berloff-2006}, using the  substitutions: $x\,{\rightarrow}\,\xi{x}$, $t\,{\rightarrow}\,(\xi^2m_1/\hbar)t$, $\psi\,\rightarrow\,\psi_{\infty}\psi$, $\varphi\,\rightarrow\,\psi_{\infty}\varphi$. Introducing new notations $\delta = m_1/m_2$, $\mu = \mu_2/\mu_1$, $\lambda = g_{12}/g_{11}$ and $\gamma = g_{22}/g_{11}$ and rearranging the terms the following system of equations can be derived
\begin{equation}
\begin{array}{c}
-2i\psi_t = \nabla^2\psi +(1-|\psi|^2)\psi - \lambda |\varphi|^2\psi , \\
-2i\varphi_t = \delta\nabla^2\varphi - \gamma|\varphi|^2\varphi - \lambda |\psi|^2\varphi + \mu\varphi ,
\end{array}
\label{system_main}
\end{equation}
with norms given by
\begin{equation}
\int |{\psi}|^2 d\vartheta = \vartheta , \,\,\, \int |{\varphi}|^2 d\vartheta = \vartheta \,\frac{N_2}{N_1} ,
\end{equation}
where $\vartheta \equiv V/\xi^3$ is a dimensionless volume.
It is easy to check that the new equation coefficients can be expressed through the primary parameters of the model in the following way
\begin{equation}
\delta = \frac{m_1}{m_2}, \,\,\,
\gamma = \frac{m_1}{m_2} \frac{l_2}{l_1}, \,\,\,
\lambda = \frac{1}{4} \left( 1 + \frac{m_1}{m_2} \right) 
\left( 1 + \frac{l_2}{l_1} \right), \,\,\,
\mu = \frac{N_2}{N_1}\frac{l_2}{l_1}\frac{m_1}{m_2} .
\end{equation}
In the dimensionless formulation the model contains only relative parameters, namely number of particles $N_2/N_1$, mass $m_2/m_1$ and scattering length $l_2/l_1$. Together with the dimensionless volume $\vartheta$ there are four model defining parameters, and three other parameters are concealed in the units. Such a formulation of the model is more universal, since it doesn't specify the type of the host superfluid, but only the relative parameters of dopants.
Introducing a dimensionless unit of energy $\varepsilon_0 =  \frac{\hbar^2}{2m_1}\xi\psi_{\infty}^2$ the Hamiltonian of the system can be written in the form
\begin{equation}
H = \int \left\{
|\nabla{\psi}|^2 + \delta |\nabla{\varphi}|^2
+\frac{1}{2} |{\psi}|^4 +\frac{1}{2}{\gamma}|\varphi|^4
+\lambda |{\psi}|^2 |{\varphi}|^2 - |{\psi}|^2 -\mu |{\varphi}|^2
\right\} d\vartheta .
\label{system_energy}
\end{equation}

In the computations the relative amount of particles $N_2/N_1 = 0.01$ is taken to study how a small amount of doping matter influences the behavior of vortices. Since dopants are considered to be larger and heavier than the host fluid particles it is assumed that $m_2 > m_1$ and $l_2 > l_1$. A range of relative parameters is scanned to investigate different types of phase-separated solutions of Eq. \ref{system_main}. The relative masses $m_2/m_1$ are varied from 5 to 45, which covers a reasonable part of the periodical table of elements. Scattering lengths of atoms are usually of the order of few angstroms, and assuming $l_2/l_1$ to vary from 1 to 10, one covers a large class of physically relevant cases. The dimensionless volume is taken as $\vartheta = 36.5^3$ which is large enough to accommodate and comfortably study a doped vortex.

All the computations are performed using the dimensionless equations. To go back to the physical domain one have to set explicitly the mass of particles $m_1$, the scattering length $l_1$ and the amount of particles $N_1$, which defines explicitly the units and the type of the superfluid. As it was already mentioned, the liquid helium is not modeled in this work, but the dopant-vortex interaction with mass and volume relations similar to the xenon-in-helium case can be considered. In this regime $m_1 = 6.646\cdot 10^{-27}$ kg and $l_1 = 2$ \AA. Taking $N_1 = 923$ a correct experimental value for the density is obtained.
Since the dimensionless volume $\vartheta$ is fixed one can derive the average density $\rho_0 = m_1\vartheta^2/N_1^2(8{\pi}l_1)^3 = 145.26$ kg/m$^3$. Healing length would be $\xi = 8{\pi}l_1N_1/\vartheta = 0.954$ {\AA}  and the unit of energy is $\varepsilon_0 = \xi\psi_{\infty}^2\hbar^2/2m_1 = 1.089\cdot 10^{-2}$ meV. Time is measured in $t_0 = 0.57$ ps. The results can be easily scaled to different BEC systems and types of fluids.

The system of Eqs. \ref{system_main} is solved numerically using the 4-th order finite difference space discretization scheme and the 4-th order Runge-Kutta method for the time propagation. The initial guess for a vortex is given by the expression
\begin{equation}
\psi(x,y,z) = \frac{x+iz}{\sqrt{x^2+z^2+\xi^2}}\psi_{\infty} .
\label{vortex_guess}
\end{equation}
Doping matter $\varphi(x,y,z)$ inside the vortex core is assumed to have a form of a filament, which repeats the cylindrical symmetry of the vortex. It is also assumed that dopants can keep the cylindrical form after decoupling (see the results).
Doping matter inside the vortex or in the superfluid bulk is initially assumed to have a gauss-like radial density distribution. The details about initial states preparation and boundary conditions can be found in the literature \cite{berloff-2000, pshenichnyuk-2016}. The imaginary time propagation technique \cite{minguzzi-2004} is used to optimize the initial states and obtain stationary characteristics before the dynamical computations start. To control the convergence, the full energy (Eq. \ref{system_energy}) is monitored and propagation proceeds until the steady state is reached.

\begin{figure*}
\centerline{\includegraphics[width=1.0\textwidth]{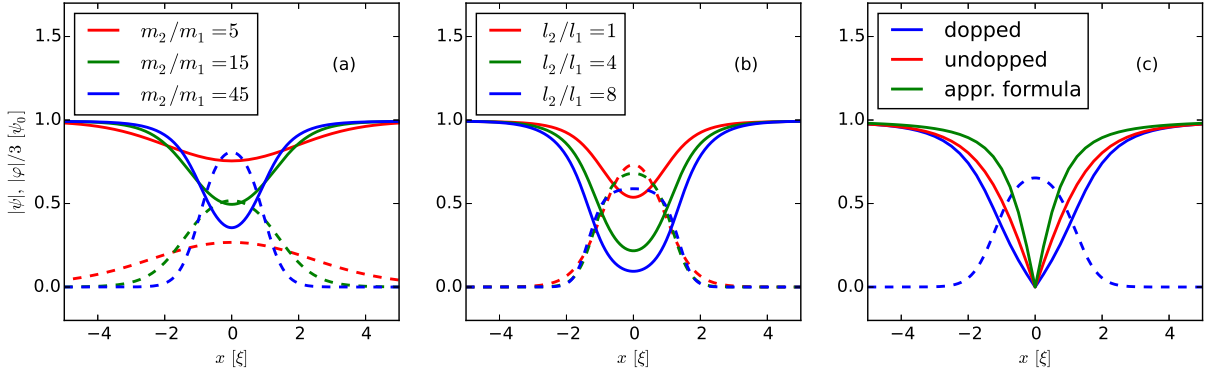}}
\caption {
A visualization of self-trapped (a,b) and vortex trapped (c) filament-like steady solutions of Eq. \ref{system_main}. Absolute values of classical fields $|\psi(x,0,0)|$ (solid lines) and $|\varphi(x,0,0)|/3$ (dashed lines) related to the fluid and dopant density distribution are plotted along $x$-axis. The panel (a) shows the regimes where $l_2/l_1=2$ and $m_2/m_1=5,15,45$ (red, green, blue). On the panel (b) $m_2/m_1=35$ is fixed and $l_2/l_1=1,4,8$ (red, green, blue). The comparison between doped and undoped vortex density profiles is shown on the panel (c) (blue and red respectively). The approximate Eq. \ref{vortex_guess} used as a vortex initial condition in the simulation is shown by the green line.
\label{fig_strapping}}
\end{figure*}

The examples of optimized steady solutions of Eq. \ref{system_main} are presented in Fig. \ref{fig_strapping}. Absolute values of classical fields $|\psi|$ and $|\varphi|$ (which are interpreted as square roots from particle's density) are plotted along $x$-axis, while $y=z=0$. The panels (a) and (b) illustrate cylindrically self-trapped filament-like doping matter placed in the bulk of the superfluid. It is located in the center of $(x,z)$-plane and alined along $y$-axis (the picture is cylindrically symmetric). The gauss-like distribution of the dopant density and corresponding minimum in the fluid density are shown. Different cases on subfigures (a) and (b) correspond to different relative masses and scattering lengths of doping particles. When the doping atoms mass approaches the mass of superfluid particles, self-trapping regime decays and no phase-separation is possible. A similar effect appears when the relative scattering length becomes small. This topic is discussed in more details in the next section.

Fig. \ref{fig_strapping} (c) demonstrates the doping substance and the fluid density distribution inside the quantum vortex core. The symmetry and orientation of this solution are the same as in the previous example. Doped and undoped vortex density profiles are plotted for the comparison. One can see that the curvature of the density slightly changes for a doped vortex, but this difference is observable only in a vicinity of a vortex core, at distances smaller than four healing lengths $\xi$. Eq. \ref{vortex_guess}, used as an initial guess for a vortex is plotted for the comparison. It overestimates the density in a vortex core slightly, but provides a good approximation at larger distances. Since no coherent (Rabi) coupling is considered in the present model, doping substance can "see" only the vortex density profile (through the term $-\lambda|\psi|^2\varphi$ in the Eq. \ref{system_main}), but not the phase of $\psi$, and the vortex works as a potential well, which accommodates the dopant. It provides additional stabilization, so that even in the regimes where no phase separation exist, vortices may isolate the second component of the mixture inside their cores.

\section{Stationary results}
\label{sec_stat}

In the experiments of Gordon and colleagues \cite{gordon-2009, moroshkin-2010}, the existence of long cylindrical metallic filaments in a superfluid is attributed to the presence of quantum vortices, while spherical metallic clusters seem to be more energetically favorable in a bulk of a vortex free liquid.
Along with the cylindrical phase separation mentioned in the previous section, spherical types of matter organization are considered here. Spherical self-trapped solutions are well studied, for example, in a framework of the electron bubble model (both in the bulk and on vortices) \cite{berloff-2000, pshenichnyuk-2016}, while less is known about cylindrical types. The question of stability of spherical and cylindrical self-trapped solutions is addressed in this section and full energies of corresponding states are compared.
The space of parameters defined in the previous section is scanned to determine the characteristics of phase-separated regimes. Such parameters as the size and binding energy, which are important to describe the matter-vortex interaction, are discussed.

A localization radius is an important parameter characterizing self-trapped spherical and cylindrical solutions. It is determined as a result of interplay between few competing terms in the full energy of the particle \cite{ghosh-2005, classen-1996}. For instance, small electrons, being placed into liquid helium, form large bubbles with radiuses of about 16 \AA. Thus, the question of size of doping particles in superfluids is not always a priori clear. A particles size is also closely connected with the binding energy to vortices, since the latter one is partially defined by the amount of rotating superfluid substituted from the vortex core by a captured impurity\cite{pshenichnyuk-2016}.
On the other hand, it is not enough to know the radius of a particle to evaluate the substitution energy, since this parameter does not take into account the degree of dopant solubility. In Fig.\ref{fig_strapping} (b) the localization changes slightly with the relative scattering length. At the same time  the superfluid density (i.e. the amount of superfluid inside the dopant) grows significantly when $l_2/l_1$ is decreased below four which corresponds to the increased solubility regime.

\begin{figure*}
\centerline{\includegraphics[width=1.0\textwidth]{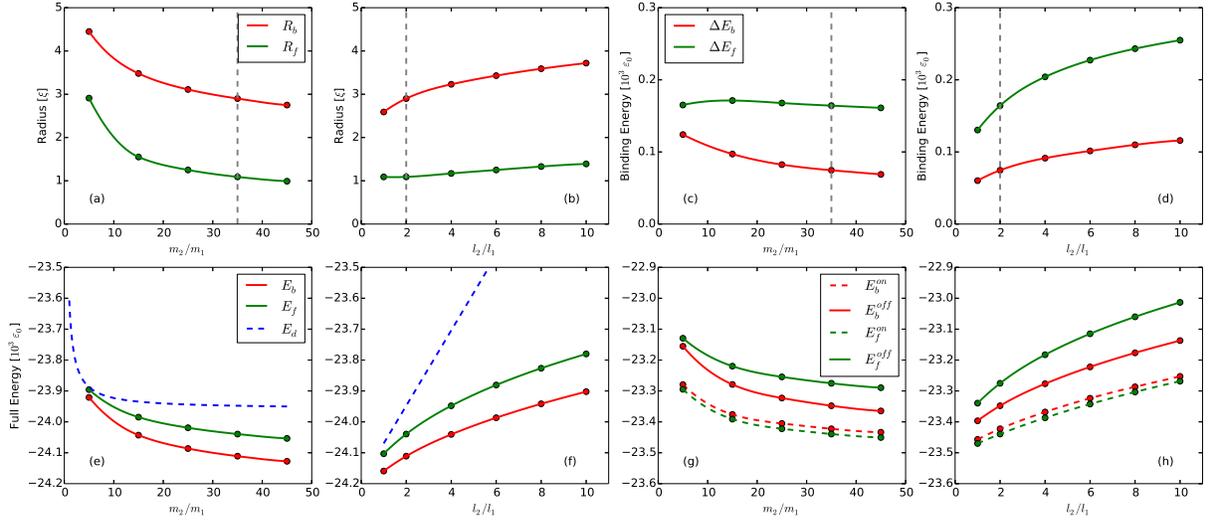}}
\caption {
Radii (a,b) and dopant-vortex binding energies (c,d) of spherical (red) and cylindrical (green) phase separated solutions of Eq. \ref{system_main} in different regimes (relative masses $m_2/m_1$ and scattering lengths $l_2/l_1$ of doping particles). All the dependencies are accompanied by the total energy plots for corresponding parameters (e),(f),(g),(h). Fully delocalized state energies are plotted with blue dashed lines on (e) and (f). On panels (g) and (h) the total energy is plotted for two cases: heavy doping particles are captured by the vortex (dashed lines) and particles are placed far from the vortex (solid lines). Vertical dashed gray lines on (a),(b),(c) and (d) approximately correspond to the xenon-in-helium regime.
\label{balvsfil}}
\end{figure*}

A radius is denoted as $R_f$ and $R_b$ for filament-like and ball-like solutions respectively. It is defined as a halfwidth of the dopant density radial distribution and computed as a function of $m_2/m_1$ and $l_2/l_1$. In addition, the full energy of corresponding states given by Eg.\ref{system_energy} in the units of $\varepsilon_0$ is computed and compared. The results are presented for spherical and cylindrical solutions in Fig.\ref{balvsfil}.
The ball radius is obviously larger than the radius of the filament (Fig.\ref{balvsfil} (a)). Both slightly decrease when the relative mass of the dopant $m_2/m_1$ is increased. When $m_2<5m_1$ the system switches to the delocalized state, where no phase separation is possible. A weakly localized state at $m_2=5m_1$ is plotted in red in Fig.\ref{fig_strapping} (a).  Even for a heavy doping matter, which is well localized, the superfluid is not completely excluded from the doping filament (it is clearly seen on Fig. \ref{fig_strapping}) and one can speak about a nonzero mutual penetration and partial solubility.
The full energy of states is presented in Fig.\ref{balvsfil} (e). Along with the localized solutions, the energy of corresponding delocalized states is plotted there. It can be easily derived from Eq.\ref{system_energy} by substituting there solutions $\psi_{\infty}$ and $\varphi_{\infty}$:
\begin{equation}
E_{deloc} = -\frac{\vartheta}{2} \left[ 1 + 
\frac{l_2}{l_1}\frac{m_1}{m_2} \left( \frac{N_2^2}{N_1^2} - \frac{1}{2}\frac{N_2}{N_1}  \right) 
-\frac{1}{2} \frac{N_2}{N_1} \left( 1 + \frac{l_2}{l_1} + \frac{m_1}{m_2} \right) \right].
\label{en_deloc}
\end{equation}
When the state becomes delocalized its energy approaches the value given by Eq.~\ref{en_deloc}. At $m_2=5m_1$ all three brunches of solutions become hardly distinguishable (Fig. \ref{balvsfil} (e)). However, filaments dissolve slightly earlier than balls, which is connected with the fact that filament-like solutions possess larger energies than balls and they are less energetically favorable.

It is shown in Fig.\ref{balvsfil} (b) how the relative scattering length $l_2/l_1$ influences the radius. Phase separated solutions exist in the whole interval of physically relevant parameters. Large values of $l_2/l_1$ correspond to the increased doping particles mutual repulsion and, thus, the larger radius. Corresponding energies are shown in Fig. \ref{balvsfil} (f). At small values of $l_2/l_1$ both spherical and cylindrical solutions approach the energy of the delocalized state and the self-trapping decays.
In contrast to the previous case (Fig.\ref{balvsfil} (a)), the dopant radius does not grow along with the increasing mutual solubility (for small values of $l_2/l_1$), but decrease slightly (Fig.\ref{balvsfil} (b)). It is also worth mentioning that in the electron bubble model, both the scattering length and mass are very small, but the solution is well localized. One important difference in our case is the existence of the repulsive term in Eq.\ref{system_main} proportional to $\gamma$. Assuming it to be zero and using simultaneously physically small values $m_2$ and $l_2$ for an electron, a bubble model with a large localization radius can be obtained. In this paper the atomic doping is considered and the parameters can not be small. For example, the relative mass and scattering length for a xenon in helium (experiments of \citet{gomez-2014}) are close to $m_2/m_1 = 35$ and $l_2/l_1=2$ values (vertical dashed gray lines on Fig. \ref{balvsfil}).

A particle-vortex binding energy is one of the most important parameters characterizing the interaction. A vortex oriented along $y$-axis is added in the center of the computational box to study binding. For the box size used, the vortex energy is $E_v = 763.1\varepsilon_0$. Being computed for helium mass and scattering length this value ($8.3$ meV) is close to the one obtained in density functional theory calculations of \citet{ancilotto-2003}, where it is appr. 0.2 meV/\AA. The computations are made for two different conformations: with the impurity placed on the vortex and impurity placed somewhere far from the vortex, then the full energies are subtracted to get the binding $\Delta{E}_{b/f}=E_{b/f}^{off}-E_{b/f}^{on}$. This procedure is applied to balls and filaments in different regimes, the results are presented in Fig. \ref{balvsfil} (c) and (d), corresponding energies are plotted in (g) and (h).

Solid red and green curves in Fig. \ref{balvsfil} (g) and (h) are very similar to curves plotted in (e) and (f). The difference is that they are shifted by a value of the vortex energy $E_v$. This approves the fact that the numerical computational box is large enough to accommodate both the impurity and the vortex far enough to exclude a significant interaction.
Dashed curves correspond to the dopant placed inside the vortex core. It is clear from the energy diagrams (g) and (h), taking into account the relative positions of red and green curves, that while the ball conformation is more energetically favorable outside the vortex, filaments are preferred inside the vortex core.

While both spherical and cylindrical solutions exist and stable in a bulk of a superfluid, the situation is slightly different inside vortex cores. It is found that balls, being captured by vortices, slowly deform themselves into filaments. This effect is less obvious for fermionic impurity models, since they do not contain the inter particle nonlinear repulsion term in the equations \cite{pshenichnyuk-2016, berloff-2000}. This term is proportional to $\gamma$ and becomes large for small $m_2$.
According to \citet{gordon-2009} the rate of guest particles clustering is higher inside the vortex core than in the bulk. In other words, particles move more freely there and the vortex core works as a "potential pipe" in helium.
Practically, during the imaginary time optimization, the fast decay of the full energy corresponds to the solution optimization, and subsequent slow decay of energy corresponds to the ball to filament transformation. Taking into account this slowness the balls on vortices configurations are considered as quasistable and placed on the energy diagram (Fig. \ref{balvsfil} (g), (h)).

As it can be seen in Fig. \ref{balvsfil} (c) and (d), according to the expectations, binding energies behave similar to the radius, growing with $l_2/l_1$ and decreasing with $m_2/m_1$. The difference appears at small masses, where a little maximum for filaments coupling $\Delta{E_f}$ is observed, despite the fact that both $E_f^{off}$ and $E_f^{on}$ decrease monotonically while $m_2/m_1$ grows. It could be related to the different terms interplay \cite{pshenichnyuk-2016} in the Hamiltonian (Eq. \ref{system_energy}).
The calculated binding energies are of the same order of magnitude as the values obtained using the density functional theory by \citet{ancilotto-2003} (order of meV) in the helium nano droplet model. In \citet{gordon-2012} binding energies are mentioned to be 0.26 - 0.86 meV per atom. 
For the range of parameters used in this paper, including both ball and filament like matter organization, bindings between $50\varepsilon_0$ and $250\varepsilon_0$ are obtained. Using the parameters for helium one gets values between 0.54 meV and 2.72 meV.
For a spherical cluster containing 9 xenon atoms (which corresponds to the chosen $N_2/N_1$ relation) the binding $\Delta{E} = 75\varepsilon_0 = 0.82$ meV is obtained. When the same amount of dopant is stretched to a filament over 34.8 {\AA} long vortex core the binding $\Delta{E} = 164\varepsilon_0 = 1.79$ meV (or 0.2 meV per atom) is obtained. It coincides with previous results \cite{pshenichnyuk-2016} calculated using a  more sophisticated model based on the nonlinear Schrodinger equation with the 7-th order nonlinearity.

The binding for filaments in the considered cases is approximately two times stronger than the binding for balls, since filaments substitute more high speed core volume of a rotating superfluid \cite{berloff-2000}. This proportion obviously strongly depends on the vortex length and the amount of doping particles. Nevertheless, decoupling of filaments does not necessary involve the whole length of the core. It is shown in Sec.~\ref{sec_dynam} how a partial filament decoupling takes place during the vortex pair reconnection.  To summarize the section it is necessary to stress, that the transition from light to heavy atomic dopants (in the regimes with a good phase separation) appears as quite smooth, with no harsh features in the main parameters of the system (Fig.~\ref{balvsfil}). It allows to concentrate the study on the influence of mass on the dynamics and comparison between the light and heavy vortices behavior.


\section{Dynamics}
\label{sec_dynam}

\begin{figure*}
\centerline{\includegraphics[width=0.9\textwidth]{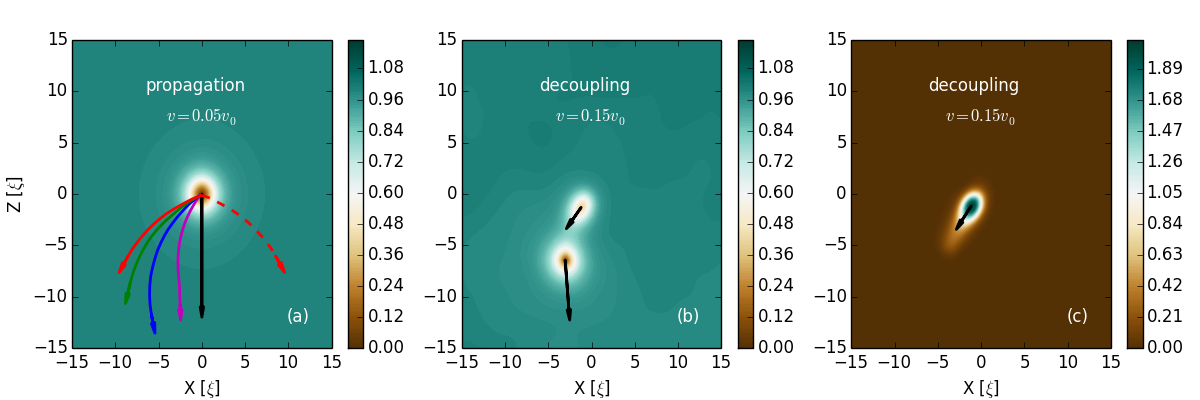}}
\caption{The panel (a) shows trajectories of a doped vortex motion plotted for different values of the relative dopant mass $m_2/m_1$. Magenta, blue, green and red curves correspond to the values 10, 20, 30 and 40 respectively. The black arrow shows the direction of the flow. The dashed red line demonstrates the vortex sign inversion result. Color coded fluid density distribution $|\psi(x,y=0,z)|$ corresponds to the initial state.
Panels (b) and (c) illustrate the decoupling phenomena, taking place at speeds higher than the critical value given by Eq. \ref{decoupling_speed}. Superfluid density $|\psi(x,y=0,z)|$ at the panel (b) shows both the vortex and the decoupled impurity (as a hole in the fluid density). Panel (c) shows the doping density $|\varphi(x,y=0,z)|$. Black arrows on (b) and (c) show the direction of velocities after the decoupling.
\label{fig_motion}}
\end{figure*}

It was shown experimentally by \citet{zmeev-2013} that vortex tangles can carry molecules through the superfluid. If the molecules are heavy enough one can expect a certain back influence on a vortex dynamics. In the regimes considered here, when $N_2/N_1=0.01$ and $m_2/m_1=40$, $m_1N_1$ and $m_2N_2$ are of the same order of magnitude, which means that the mass of the dopant is of the same order as the total mass of particles forming a vortex. The dynamical behavior of such heavy composite formations is analyzed in this section.

The computation starts from the initial state, considered in the previous section, where a doping filament trapped by the vortex is located in the center of the computational box (2D cross-section is depicted in Fig. \ref{fig_motion}). Then Eq.~\ref{system_main} is solved in the moving reference frame. The transformation is realized by applying the following modification:
$\psi_t \rightarrow \psi_t - (\mathbf{v}_{un}\cdot\nabla)\psi$,
where $\mathbf{v}_{un}$ is a speed of a uniform fluid flow, measured in the units of $v_0=\hbar/{\xi}m_1$. To avoid high speed hydrodynamic phenomena such as a vortex rings generation \cite{pshenichnyuk-2015} the speed value is taken well below criticality such that $v_{un}^x=v_{un}^y=0$ and $v_{un}^z = 0.05$.
Such a uniform superfluid flow, obviously, provides no viscous drag to the phase separated doping matter, but it moves the vortex which drags the bonded dopant.

The resulting trajectories of doped vortices with different mass parameters are presented in Fig. \ref{fig_motion} (a) (to exclude the influence of boundaries the calculations were repeated in a larger box). The higher the doping mass is, the larger is the deviation from the straight line propagation.
The physical nature of the force, pushing the vortex in the direction perpendicular to the propagation direction, can be understood using the Bernoulli's principle
\begin{equation}
\frac{\rho{v^2}}{2} + p = const ,
\label{bern_conserv}
\end{equation}
where $p$, $\rho$ and $v$ are pressure, density and velocity of a fluid at a certain point in space.
As a result of summation of a vortex velocity field and the uniform flow field, absolute values of fluid velocities at different sides of the dopant become different, which causes the pressure difference and the appearance of the corresponding force. The direction of the force can be inverted by changing the sign of $v_{un}^z$ or the winding number of the vortex. The latter case is demonstrated in Fig. \ref{fig_motion} (a) by red dashed line.
This force is similar to the classical Magnus force, which acts on a rotating cylinder moving in a liquid or gas. The difference with respect to the classical case is that the doping itself does not rotate at all, but the rotational motion of the surrounding superfluid is caused by the quantum vortex attached to the body.

The Magnus force disappears when the doping filament moves together with the flow and their relative velocity is zero. The pressure in this case is the same everywhere around the filament. If the dopant mass is small it can be quickly accelerated by the flow, but it takes more time to accelerate a heavy filament. In the second case Magnus force acts longer and causes larger deviations. At the same time one can say that it is easier for a heavy doping filament to drag the vortex aside. In the regimes considered in this paper vortex pinning (when the dopant prevents vortices from moving) was not observed. 

To evaluate the force mathematically, one may start from the general expression for the Bernoulli force and take into account the conservation law (Eq. \ref{bern_conserv}) to write \cite{sergeev-2006}
\begin{equation}
\mathbf{F} = \int\limits_S p(\mathbf{r}) \mathbf{n}(\mathbf{r}) ds = 
\int\limits_S \frac{\rho v^2}{2} \mathbf{n}(\mathbf{r}) ds ,
\end{equation}
where the integral is taken over the surface of a body, embedded into the fluid, and $\mathbf{n}(\mathbf{r})$ is a unit vector normal to the surface $S$.
The body is assumed to be a cylinder with a radius $R$ and length $L$. Using the symmetry and introducing the polar coordinates one may write
\begin{equation}
\mathbf{F} = \frac{L}{2} \int\limits_{0}^{2\pi} \rho(R,\alpha) v^2(R,\alpha) \binom{\cos\alpha}{\sin\alpha} R \, d\alpha
\label{bern_force2}
\end{equation}
(there is no force component acting along $y$-axis and we consider only $x$ and $z$ components).
Assuming the classical field in the form $\psi = |\psi| e^{i\phi}$, the fluid  density and velocity read: $\rho = |\psi|^2 m_1$, $\mathbf{v}_s = \frac{\hbar}{m_1}\nabla\phi$. In Eq.~\ref{bern_force2}, $v$ is an absolute value of a sum of the vortex velocity and the uniform flow velocity fields, i.e. $v = |\mathbf{v}_{v} + \mathbf{v}_{f}|$.
To compute $\mathbf{v}_{v}$ one may use an approximate expression for a vortex order parameter (Eq. \ref{vortex_guess}),
and neglect the disturbance caused by the presence of the dopant (Fig. \ref{fig_strapping}). In this case 
\begin{equation}
|\psi|^2 =  \frac{\psi^2_{\infty} R^2}{R^2 + \xi^2},
\end{equation}
\begin{equation}
\nabla\phi = \frac{1}{R^2} \binom{-z}{+x}.
\end{equation}
Assuming $\mathbf{v}_{f}$ oriented in the negative direction of $z$-axis, one gets
\begin{equation}
v^2 = v^2_{v} + v^2_{f} + 2 \mathbf{v}_{v} \cdot \mathbf{v}_{f}
= \frac{\hbar^2}{m_1^2 R^2} + v_{f}^2 - \frac{2\hbar{v_{f}}}{m_1} \frac{x}{R^2}
\end{equation}
and
\begin{equation}
\frac{\rho v^2}{2} = \frac{\psi_{\infty}^2}{R^2 + \xi^2}
\left[  \frac{\hbar^2}{2m_1} + \frac{m_1v_{f}^2R^2}{2} - \hbar{v_{f}}x  \right] .
\end{equation}
In the last equation first two terms in the brackets do not depend on $\alpha$ and being substituted into the integral in Eq. \ref{bern_force2} vanish. The expression for the force reads
\begin{equation}
\mathbf{F} = -LR^2 \frac{\psi^2_{\infty}\hbar{v_{f}}}{R^2+\xi^2} \int\limits_{0}^{2\pi}
 \cos\alpha \binom{\cos\alpha}{\sin\alpha} d\alpha ,
\end{equation}
where only the $x$-component survives the integration
\begin{equation}
F_x = -L\hbar v_{f} \pi \frac{\psi^2_{\infty}R^2}{R^2+\xi^2}.
\end{equation}
Having in mind the classical situation of an ideal liquid flowing around a cylindrical obstacle it is assumed that $v_f\approx 2v_{un}v_0$.

The dynamics of doped vortices in the presented examples is defined by an interplay of two forces: the vortex drag acting on doping filaments and the Magnus force. Such a coupled vortex-dopant dynamics has its limits, defined by the binding energy. The decoupling happens when the kinetic energy of the relative movement exceeds the value $\Delta{E}$ (see Fig.~\ref{balvsfil} (c) and (d)). Mathematically it could be formulated as $N_2m_2v_{dec}^2/2=\Delta{E}\,\varepsilon_0/v_0^2$, or, expressing everything through the  primary parameters
\begin{equation}
v_{dec} = \sqrt{\frac{\Delta{E}}{\vartheta}\frac{m_1N_1}{m_2N_2}}.
\label{decoupling_speed}
\end{equation}
Using the value of $\Delta{E}$ presented in Fig. \ref{balvsfil} for heavy dopants with $m_2/m_1=40$ one gets $v_{dec}^{40} = 0.1$. This value is larger for relatively light dopants when, for instance, $m_2/m_1=10$ and $v_{dec}^{10} = 0.2$. It is easier to "shake off" heavy dopants, since the decoupling happens at lower velocities.

The decoupling process for $m_2/m_1=40$ and $v_{un}=0.15$ is illustrated in Fig. \ref{fig_motion} (b) and (c), where 2D cross-sections for $\psi$ and $\varphi$ absolute values are shown. The vortex and the doping filament completely separate at $t=60$. Moving directions after the separation are shown by the arrows. According to Eq. \ref{decoupling_speed}, the decoupling appears at higher speeds for smaller masses, which is approved by the simulations (not shown). In Fig. \ref{fig_motion} (c) the decoupled doping density is slightly stretched along the direction of movement.  For light regimes a small density split effect is observed, when a certain portion of the dopant remains attached to the vortex. This effect is similar to the one mentioned in \citet{pshenichnyuk-2016}.

\begin{figure*}
\centerline{\includegraphics[width=1.0\textwidth]{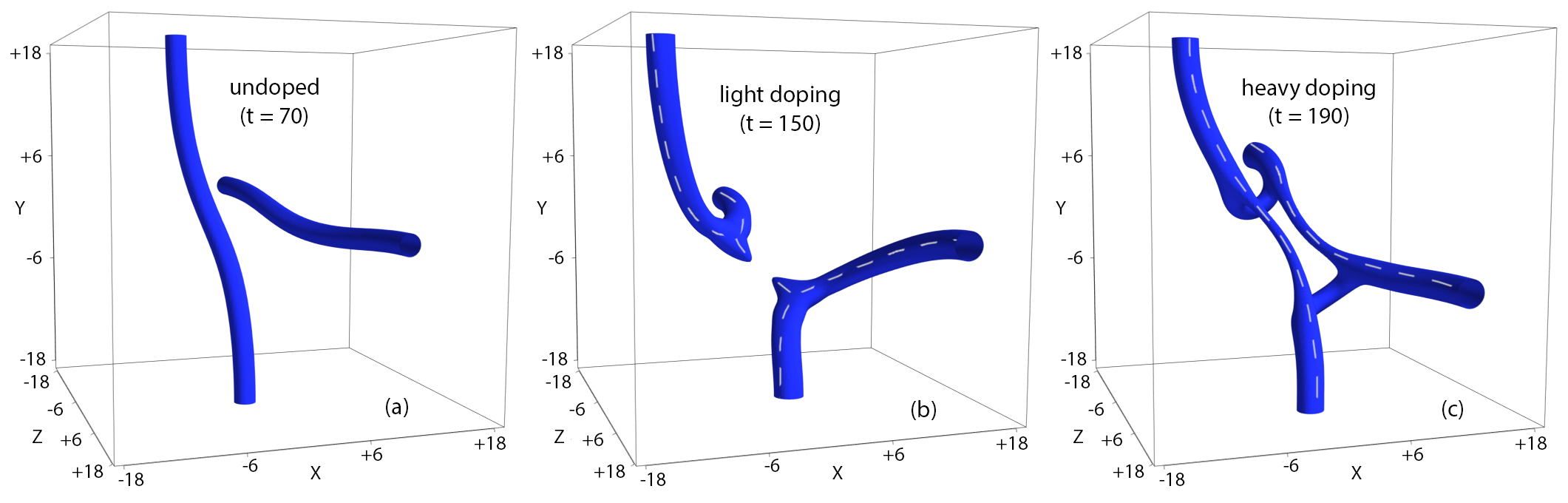}}
\caption{
The reconnection of two orthogonal quantum vortices initially separated by a distance $10\xi$. The superfluid density isosurface $|\psi(x,y,z)|=const$ is plotted. The panel (a) shows an early stage of the evolution for undoped vortices, which looks similar also for doped cases. Panels (b) and (c) show the result of the reconnection for light and heavy vortices respectively. Thin dashed lines mark the layout of doping filaments.
\label{fig_reconnection}}
\end{figure*}

In rotating helium nanodroplets quantum vortices form an arranged structure called the vortex lattice. In the experiments of Gordon and colleagues, quite opposite, vortex filaments form a chaotic turbulent state. It is known that a vortex tangle evolves through multiple vortex reconnection events. It is an example of a fundamental process which can be used to illustrates the role of the mass of doping filaments. The reconnection goes differently for light and heavy vortices (see Fig. \ref{fig_reconnection} (b) and (c) ). At the early stages of the evolution two perpendicular vortices, separated by a distance $10\xi$, start to bend and behave similar to the well studied undoped case (see Fig. \ref{fig_reconnection} (a)). The process for the light vortex then goes slightly faster: the core interaction starts at $t=70$, while in heavy case it happens at $t=120$. Comparing the results of reconnection (Fig.\ref{fig_reconnection} (b) and (c)) one may see that in the first case doping filaments inside the cores (shown by dashed gray lines) reconnect following the motion of host vortices. In the second case the rigid frame, produced by heavy doping filaments, remains almost unchanged, while vortices reconnect and a qualitatively different picture is observed.
In the heavy doping case a partial decoupling takes place, when only a certain fraction of the the doping filament is released from the vortex core.
Similar processes are expected to be common during the laser ablation of heavy metals in superfluids. Theoretical modeling should help to acquire a better understanding of mechanisms leading to the production of metallic nanowires in corresponding experiments.

\section{Conclusion}
\label{sec_conclusion}

In this paper the general questions of a doped quantum vortex properties and behavior are addressed. 3D mathematical model based on a system of nonlinear matter field equations is used to describe the interaction of quantum vortices with doping matter. Phase separated solutions (where the doping substance is separated from the rest of the fluid) of different geometries are investigated. Stability of solutions and possible vortex induced geometry transformations are discussed. The emphasis is made on the influence of dopant mass on the vortex behavior. Sizes of doping matter fractions as well as their binding energy to vortices in a wide range of physically relevant parameters are investigated. 

It is shown that the motion of doped vortices is influenced by the Magnus force which acts on the doping particles and drags the vortex in the direction perpendicular to the direction of the flow. The role of the dopant mass in this effect is discussed and the analytical expression for the force is derived. The computed vortex-dopant binding energies are used to formulate the decoupling criteria and simulate the corresponding process.
The reconnection process is simulated to demonstrate the difference between the light and heavy vortex behavior. In the first case the dopant readily follows the vortices during the reconnection. In the second case vortices partially decouple from the heavy dopant frame which keeps its original topology during the simulation time. 
To conclude the obtained results it should be stressed that heavy dopants can not be considered as just a passive visualization tool in superfluid experiments, since they can influence significantly the vortex behavior.

\section*{Acknowledgment}

Fruitful discussions with Natalia Berloff and Andrey Vilesov are greatly acknowledged.

\bibliography{paper_heavycores}

\end{document}